\documentclass{article} 
\usepackage{epsf}
\usepackage{graphicx}
\usepackage{threeparttable}
\usepackage{amsmath} 
 
\begin{document} 
 
\def \be {\begin{equation}} 
\def \ee {\end{equation}} 
\def \bea {\begin{eqnarray}} 
\def \eea {\end{eqnarray}} 
\def \bse {\begin{subequations}} 
\def \ese {\end{subequations}} 
\def \bde {\begin{description}} 
\def \ede {\end{description}} 
\def \nn {\nonumber} 
\def \spa {$\;$} 
\def \la {\langle} 
\def \ra {\rangle} 
\def \R {{\bf R}} 
\def \C {{\bf C}} 
\def \Z {{\bf Z}} 
\def \del {\partial} 
\def \dels {\partial\kern-.5em / \kern.5em} 
\def \As {{A\kern-.5em / \kern.5em}} 
\def \Ds {D\kern-.7em / \kern.5em} 
\def \Psib {{\bar \Psi}}
\def\e{\,{\rm e}}

\def \rh {\kappa} 
\def \a {\alpha} 
\def \b {\beta} 
\def \dag {\dagger} 
\def \g {\gamma} 
\def \G {\Gamma} 
\def \d {\delta} 
\def \D {\Delta}
\def \eps {\epsilon} 
\def \m {\mu} 
\def \n {\nu} 
\def \k {\kappa} 
\def \lam {\lambda} 
\def \Lam {\Lambda} 
\def \s {\sigma} 
\def \r {\rho} 
\def \om {\omega} 
\def \Om {\Omega} 
\def \one {{\bf 1}} 
\def \th {\theta} 
\def \Th {\Theta} 
\def \Chi {\chi} 
\def \t {\tau} 
\def \ve {\varepsilon} 
\def \II {I\hspace{-.1em}I\hspace{.1em}} 
\def \IIA {\mbox{\II A\hspace{.2em}}} 
\def \IIB {\mbox{\II B\hspace{.2em}}}

\def\np#1{{\sl Nucl.~Phys.~\bf B#1}}
\def\pl#1{{\sl Phys.~Lett.~\bf B#1}}
\def\pr#1{{\sl Phys.~Rev.~\bf D#1}}
\def\prl#1{{\sl Phys.~Rev. Lett.~\bf #1}}
\def\cm#1{{\sl Comm.~Math.~Phys.~\bf #1}}
\def\mpl#1{{\sl Mod.~Phys.~Lett.~\bf A#1}}
\def\cpc#1{{\sl Comp.~Phys.~Comm.~\bf #1}}
\def\anp#1{{\sl Ann.~Phys.~(NY) \bf #1}}
\def\rmp#1{{\sl Rev.~Mod.~Phys. \bf #1}}
\def\cqg#1{{\sl Class.~Quant.~Grav.~\bf #1}}
 
\newcommand{\ho}[1]{$\, ^{#1}$} 
\newcommand{\hoch}[1]{$\, ^{#1}$} 
 
%%%%%% These are only for this paper %%%%%% 
 
\def \H {{\cal H}} 
\def \MSUSY {M_{\mbox{\tiny SUSY}}}. 
\def \S {{\cal S}} 
\def \ha {\hat{a}} 
\def \u {u} 
\def \z {\om} 
\def \bigdot {\mbox{\Large .}}
 
%\begin{titlepage} 
%\catcode`\@=11 
%\catcode`\@=12 
%\twocolumn[\hsize\textwidth\columnwidth\hsize\csname% 
%@twocolumnfalse\endcsname 
 
%\draft 
\begin{center} 
\hfill hep-th\\ 
\vskip .5in 
 
\textbf{\large 
Second Order Perturbative Calculation of Quasinormal Modes of Schwarzschild Black Holes 
} 
 
\vskip .5in 
{\large
Hsien-chung Kao 
\vskip 15pt 
 
{
{Department of Mathematics, University of Durham, Durham, DH1 3LE, UK.} \footnote{on leave from National Taiwan Normal University.} }\\ 
{Department of Physics, National Taiwan Normal University, Taipei, Taiwan 116.}\\ 
}
  
\vskip .2in 
\sffamily{ 
hckao@phy.ntnu.edu.tw}
 
\vspace{60pt} 
%\maketitle 
\end{center} 
\begin{abstract} 
  
We analytically calculate to second order the correction to the asymptotic form of quasinormal frequencies of four dimensional Schwarzschild black holes based on the monodromy analysis proposed by Motl and Neitzke.  Our results are in good agreement with those obtained from numerical calculation. 
  
\end{abstract} 
%\pacs{ PACS numbers: 04.50.+h, 04.70.Bw, 11.25.-w, 11.27.+d} 
%\end{titlepage} 
%\begin{narrowtext} 
 
\setcounter{footnote}{0} 
\newpage 
 
\section{Introduction} 
 
Quasinormal modes (QNMs) were originally observed in considering the scattering or emission of gravitational waves by Schwarzschild black holes \cite{Vish}. It was found that a characteristic damped oscillation, which only depends on the black hole mass, dominated the time evolution in a certain period of time.  Since then QNMs have been investigated extensively both analytically and numerically. For a general review and classification, see Refs. \cite{NollertReview,Schiappa}. From numerical studies, an asymptotic formula for quasinormal frequencies of Schwarzschild black holes was obtained \cite{Nollert}:
\be
2GM \om_n \approx 0.0874247 + {1\over 2}\left(n- {1\over 2} \right)i + {\sc O}[n^{-1/2}]. 
\ee 
The real part in the above formula was later postulated to be ${1\over 4\pi}\ln 3$ \cite{Hod} based on a discrete area spectrum of quantum black holes proposed in Ref. \cite{Bekenstein}.  This was confirmed later by Motl and Neitzke \cite{Motl}.  The recent surge of interest in the QNMs derived from its possible application in determining the Immirzi parameter in loop quantum gravity\cite{Dreyer}.  The numerical value $\ln3$ in the real part of the asymptotic quasinormal frequencies in Schwarzschild black holes was at first taken as a hint that the relevant gauge group in loop quantum gravity is $SO(3)$ instead of the commonly believed $SU(2)$. However, as shown in Ref. \cite{Motl}, the value $\ln3$ is not universal and one should take the argument with a grain of salt. 

Another interesting application of QNMs was pointed out by Horowitz and Hubeny in their study of a scalar field in the background of a Schwarzschild anti-de Sitter  black hole \cite{Horowitz}. According to AdS/CFT correspondence, a large black hole in AdS spacetime corresponds to a thermal state in CFT \cite{AdSCFT}. They argued the decay of the scalar field corresponds to the decay of a perturbation of this state. In the BTZ black hole, a one-to-one correspondence was found between the QNMs in the bulk and the poles of the retarded correlation function in the dual conformal field theory on the boundary \cite{Birmingham}. The idea of dS/CFT correspondence has also been proposed and formulated \cite{dSCFT}.  Since there is a cosmological horizon in de Sitter spacetime, QNMs may also be defined in principle.  Similar studies of QNMs have also been carried out in de Sitter spacetime trying to lent support for such correspondence \cite{Abdalla}. However, the situation there is more subtle and it seems QNMs only exist in odd dimensions \cite{Schiappa}. Therefore, it is not clear whether such correspondence makes sense in even dimensions, and further study is necessary.

\section{Perturbative calculation of the asymptotic form of quasinormal frequencies}

In Ref. \cite{Brink}, the author calculated the first order correction to the asymptotic form of quasinormal frequencies of a Schwarzschild black hole using a WKB analysis. The result was extended to include the scalar field case using the monodromy analysis developed by Motl and Neitzke \cite{Siopsis}. The agreement with numerical results is excellent. We will begin with a brief review of their method which made systematic expansion more accessible.  In a background spacetime described by a metric $g_{\m\n}$, a massless scalar $\Phi$ satisfies the following Klein-Gordon equation:
\begin{eqnarray}
{1 \over \sqrt{-g}}\partial_{\mu}\left\{g^{\mu \nu}\sqrt{-g}\partial_{\nu}\Phi \right\}=0.
\end{eqnarray}
For four dimensional Schwarzschild black holes, the metric is given by 
\begin{displaymath}
{ds{^2}=-f(r) dt{^2}+f(r)^{-1} dr{^2}+r{^2}d\Omega{^2}},
\end{displaymath}
with $f(r) = (1-{r_0 \over r})$ and $r_0 = 2GM.$ 
Let 
\be
\Phi(r,t,\Omega)= r\, \phi(r) Y_{lm}(\Om) \e^{i\om t}. \label{Fourier}
\ee
$\phi(r)$ now satisfies the following equation:
\begin{eqnarray}
-f(r) {d \over dr} \left[ f(r) {d \phi\over dr} \right] + V(r) \phi = \om^2 \phi, \label{AD}
\end{eqnarray} 
with 
\begin{displaymath}
V(r)=(1-{r_0 \over r})\left[{l(l+1) \over r{^2}} + {r_0 \over r{^3}}\right].
\end{displaymath}
By a simple modification in the potential $V(r)$ \cite{NollertReview},
\begin{eqnarray}
V(r)=(1-{r_0 \over r})\left[{l(l+1) \over r{^2}} + {(1-j^2)r_0 \over r{^3}}\right],\label{AE}
\end{eqnarray}
the previous equation can also describes linearized perturbation of the metric or an electromagnetic test fields.  Here, $j=0,1,2$ which is the spin of the relevant field.  They can also be classified as the tensor, vector, and scalar types of perturbation to the background Schwarzschild metric using the master equations derived by Ishibashi and Kodama \cite{Ishibashi}. 
Introducing the tortoise coordinate:
\begin{displaymath}
x(r)=r+ r_0 \ln (r/r_0-1 ), 
\end{displaymath}
one obtain a Schrodinger-like equation 
\begin{eqnarray}
\left\{ -{d^2 \over dx^2} + V[r(x)] \right\} \phi= \om^2 \phi.
\end{eqnarray}

Because of our convention in eq (\ref{Fourier}), QNMs are defined through the following out-going wave boundary condition:
\begin{eqnarray}\label{AL2}
\phi(x) \sim \left\{
\begin{array}{ll} 
\e{^{i\omega x}} & \mbox{as x} \rightarrow -\infty \mbox{ (horizon)}, \\
\e{^{-i\omega x}} & \mbox{as x} \rightarrow \infty \quad \mbox{(spatial infinity)},
\end{array}\right.
\end{eqnarray} 
assuming ${\rm Re}\, \om >0$. 
Define  
\be
f(x) = \e^{i\om x}\, \phi \sim  \left\{
\begin{array}{ll} 
\e{^{2i\omega x}} & \mbox{as x} \rightarrow -\infty, \\
1                 & \mbox{as x} \rightarrow \infty.
\end{array}\right.
\ee
According to Ref. \cite{Motl}, the boundary condition at the horizon translates to the monodromy of $f(x)$ around it
\be
{\cal M}(r_0) = \e^{4\pi \om r_0}.
\ee 

The same monodromy can also be accounted for by those around $r=0$ and $r = \infty$, and it has  been shown that only the former one is non-trivial.  To find the monodromy around $r=0$, one need to introduce the complex coordinate variable
\be
z = \om(x-i\pi r_0) = \om[ r+r_0\ln(1-r/r_0)],
\ee
which is vanishing at the black hole singularity $r=0$.
In the limit $|r/r_0| \ll 1$, the potential can be expanded as a series in $\sqrt{z/(\om r_0)}$:
\be
V(z)= - {\om^2 (1-j^2) \over 4z^2}  + {3 l(l+1)+1-j^2 \over 6 \sqrt{2} (-\om r_0)^{1/2} z^{3/2}} - {3 l(l+1)+1-j^2 \over 36 \sqrt{2} (-\om r_0)^{3/2} z^{1/2}} + \dots.
\ee
Note that the third term in the above expression is of order $(-\om r_0)^{-3/2}$ and would not contribute until we consider third order perturbation. 
To second order in perturbation theory, the wavefunction can be expanded as
\be
\phi = \phi^{(0)} + {1\over \sqrt{-\om r_0}} \phi^{(1)} + {1\over -\om r_0} \phi^{(2)} + O(\om^{-3/2}).
\ee
The zeroth, first and second order equations are given by
\bea
&\;& {d\phi^{(0)} \over d z^2} + \left({1-j^2 \over 4z^2} +1 \right) \phi^{(0)}=0; \\
&\;& {d\phi^{(1)} \over d z^2} + \left({1-j^2 \over 4z^2} +1 \right) \phi^{(1)}= \sqrt{-\om r_0}\, \d V(z)\, \phi^{(0)}; \\
&\;& {d\phi^{(2)} \over d z^2} + \left({1-j^2 \over 4z^2} +1 \right) \phi^{(2)}= \sqrt{-\om r_0}\, \d V(z)\, \phi^{(1)},
\eea
respectively.  Here,
\be
\d V(z) = {3 l(l+1)+1-j^2 \over 6 \sqrt{2} (-\om r_0)^{1/2} z^{3/2}}.
\ee
Define $\phi_{\pm}^{(0)}(z)$ to be the two linearly independent solutions to the zeroth order equation
\be
\phi_{\pm}^{(0)}(z) = \sqrt{\pi z\over 2} J_{\pm j/2}(z). \label{zeroth}
\ee
In the asymptotic region $z\gg 1$
\be
\phi_{\pm}^{(0)}(z) \approx \cos[z - \pi(1\pm j)/4].
\ee
It has been shown by Musiri and Siopsis that $\phi_{\pm}^{(1)}$ can be expressed in terms of $\phi_{\pm}^{(0)}$
\bea
&\;& \hskip -1cm \phi_{+}^{(1)}(z) 
= C \phi_{+}^{(0)}(z) \int_{0}^{z} dz_1\, \d V(z_1)\, \phi_{-}^{(0)}(z_1)\, \phi_{+}^{(0)}(z_1)  
- C \phi_{-}^{(0)}(z) \int_{0}^{z} dz_1\, \d V(z_1)\, \phi_{+}^{(0)}(z_1)\, \phi_{+}^{(0)}(z_1); \\
&\;& \hskip -1cm \phi_{-}^{(1)}(z) 
= C \phi_{+}^{(0)}(z) \int_{0}^{z} dz_1\, \d V(z_1)\, \phi_{-}^{(0)}(z_1)\, \phi_{-}^{(0)}(z_1)  
- C \phi_{-}^{(0)}(z) \int_{0}^{z} dz_1\, \d V(z_1)\, \phi_{+}^{(0)}(z_1)\, \phi_{-}^{(0)}(z_1).
\eea
where $C = \sqrt{-\om r_0}/\sin(\pi j/2)$ \cite{Siopsis}.
Similarly, $\phi_{\pm}^{(2)}$ can in turn be expressed in terms of $\phi_{\pm}^{(1)}$
\bea
&\;& \hskip -1cm \phi_{+}^{(2)}(z) 
= C \phi_{+}^{(0)}(z) \int_{0}^{z} dz_2\, \d V(z_2)\, \phi_{-}^{(0)}(z_2)\, \phi_{+}^{(1)}(z_2)  
- C \phi_{-}^{(0)}(z) \int_{0}^{z} dz_2\, \d V(z_2)\, \phi_{+}^{(0)}(z_2)\, \phi_{+}^{(1)}(z_2); \\
&\;& \hskip -1cm \phi_{-}^{(2)}(z) 
= C \phi_{+}^{(0)}(z) \int_{0}^{z} dz_2\, \d V(z_2)\, \phi_{-}^{(0)}(z_2)\, \phi_{-}^{(1)}(z_2)  
- C \phi_{-}^{(0)}(z) \int_{0}^{z} dz_2\, \d V(z_2)\, \phi_{+}^{(0)}(z_2)\, \phi_{-}^{(1)}(z_2).
\eea
In the limit, $z\to \infty$,
\bea
&\;& \hskip -1cm  \phi_{\pm}^{(1)}(z) 
=  c_{-\pm}\, \phi_{+}^{(0)}(z)  - c_{+\pm}\, \phi_{-}^{(0)}(z) ; \\
&\;& \hskip -1cm  \phi_{\pm}^{(2)}(z) 
=  d_{-\pm}\, \phi_{+}^{(0)}(z)  - d_{+\pm}\, \phi_{-}^{(0)}(z) .
\eea
Here, 
\bea
&\;& \hskip -2cm
c_{\pm\pm} = C \int_{0}^{\infty} dz_1\, \d V(z_1)\, \phi_{\pm}^{(0)}(z_1)\, \phi_{\pm}^{(0)}(z_1); \\
&\;& \hskip -2cm 
d_{\pm\pm} = C^2 \int_{0}^{\infty} \,dz_2  \int_{0}^{z_1} \,dz_1 \d V(z_2)\,\d V(z_1)\, \phi_{\pm}^{(0)}(z_2)\left[ \phi_{+}^{(0)}(z_2) \phi_{-}^{(0)}(z_1) - \phi_{-}^{(0)}(z_2) \phi_{+}^{(0)}(z_1) \right]  \phi_{\pm}^{(0)}(z_1). 
\eea
Notice that $\phi_{\pm}^{(0)}$ defined in eq (\ref{zeroth}) are in fact linearly dependent to each other when $j$ is an even integer.  As a result, each of these coefficients is divergent by itself in these cases.  It is reassuring to see that all the divergent pieces cancel among themselves so that physically interested  quantities do have a smooth limit when $j$ is an even integer. 
In zeroth order, the combination 
\be
\phi^{(0)}(z) = \phi_{+}^{(0)}(z) - \e^{-i\pi(j/2)} \phi_{-}^{(0)}(z) \sim \e^{-iz}
\ee
in the asymptotic region $z\gg 1.$
This can be extended to second order 
\bea
&\;& \hskip -2cm  \phi(z) = \left[\phi_{+}^{(0)}(z) +  {1\over \sqrt{-\om r_0}} \phi_{+}^{(1)}(z) + {1\over -\om r_0} \phi_{+}^{(2)}(z) \right]  \nonumber \\
&\;& \hskip -1.0cm - \e^{-i\pi(j/2)} \left[1 - {\xi \over \sqrt{-\om r_0}}  - {\zeta \over -\om r_0}  \right] \left[\phi_{-}^{(0)}(z) +  {1\over \sqrt{-\om r_0}} \phi_{-}^{(1)}(z) + {1\over -\om r_0} \phi_{-}^{(2)}(z) \right], \label{phi}
\eea
by introducing two parameters $\xi$ and $\zeta$.  Naturally, they are determined by the condition that the coefficient of the $\e^{iz}$ term is vanishing when $z\to \infty$:
\bea
&\;& \hskip -2cm  \xi =  \xi_{+} + \xi_{-}; \\
&\;& \hskip -2cm  \zeta = - \xi \xi_{-} + d_{++}\e^{i\pi j/2} - d_{+-} + d_{--}\e^{-i\pi j/2} - d_{-+},
\eea
where 
\bea
&\;& \xi_{+} = c_{++}\e^{i\pi j/2} - c_{+-}, \quad  \xi_{-} = c_{--}\e^{-i\pi j/2} - c_{-+}. 
\eea
Substitute the above result back to eq (\ref{phi}), we have
\bea
&\;& \hskip -2cm  \phi(z) = i \e^{i\pi(1-j)/4} \sin(\pi j/2) \e^{-iz} \left\{ 1 - {\xi_{-} \over \sqrt{-\om r_0}}  + {\xi(\xi_{m} +c_{+-}) - d_{--} \e^{-i\pi j/2} + d_{-+}\over -\om r_0} \right\}, \label{a2}
\eea
where the identity $c_{-+} = c_{+-}$ has been used to simplify the expression.

When going around the black hole singularity by $3\pi$, $\phi_{\pm}^{(1)}$ and $\phi_{\pm}^{(2)}$ both pick up an extra phase:
\bea
&\; \phi_{\pm}^{(1)}(\e^{3i\pi}z) = \e^{3i\pi(2\pm j)/2}\phi_{\pm}^{(1)}(-z); \\
&\; \phi_{\pm}^{(2)}(\e^{3i\pi}z) = \e^{3i\pi(3\pm j)/2}\phi_{\pm}^{(2)}(-z).
\eea
Consequently,
\bea
&\;& \hskip -3.5cm  \phi(\e^{3i\pi} z) = \e^{3i\pi(1+j)/2} \left[\phi_{+}^{(0)}(-z) - i  {1\over \sqrt{-\om r_0}} \phi_{+}^{(1)}(-z) - {1\over -\om r_0} \phi_{+}^{(2)}(-z) \right]  \nonumber \\
&\;& \hskip -1.9cm - \e^{-i\pi(j/2)} \left[1 - {\xi \over \sqrt{-\om r_0}} - {\zeta \over -\om r_0} \right] \nonumber \\
&\;& \hskip -1.7cm \e^{3i\pi(1-j)/2} \left[\phi_{-}^{(0)}(-z) -i  {1\over \sqrt{-\om r_0}} \phi_{-}^{(1)}(-z) - {1\over -\om r_0} \phi_{-}^{(2)}(-z) \right].
\eea
To second order,
\bea
&\;& \hskip -2cm  \phi(\e^{3i\pi}z) = -i \e^{i\pi(1-j)/4} \sin(3\pi j/2) \e^{-iz}  \nonumber \\
&\;& \hskip -0.3cm \left\{ 1 +  {(1+ ie^{3i\pi j})\xi_{+} + (1+i)\xi_{-} \over \sqrt{-\om r_0} (-1+ \e^{i3\pi j})} \right. \nonumber \\
&\;& \hskip -0.1cm \left. + { -(1+i)\xi\, \xi_{-} + [(1 + \e^{i3\pi j})(d_{++}\e^{i\pi j/2} - d_{-+}) + 2d_{--}\e^{-i\pi j/2} - 2d_{+-} ] \over -\om r_0 (-1+ \e^{i3\pi j})}  \right\} + \ldots , \label{a3}
\eea
where the term $\e^{iz}$ is not relevant for our calculation and has been neglected. Taking the ratio between the coefficients of the term $\e^{-iz}$ in eqs (\ref{a3}) and (\ref{a2}), we obtain the monodromy to second order:
\bea
&\;& \hskip -2cm {\cal M}(r_0) = -[1 + 2\cos(j\pi)] 
\left\{ 1 + {\D_1 \over \sqrt{-\om r_0} } + { \D_{2c} + \D_{2d} \over  -\om r_0 } \right\}.
\eea
Here,
\bea
&\;& \D_1 = {(1+ ie^{3i\pi j})\xi_{+} + (i + e^{3i\pi j})\xi_{-} \over (-1+ \e^{i3\pi j})}; \\
&\;& \D_{2c} = - (1-i)\xi_{+}\, \xi_{-} - \xi c_{+-}; \\
&\;& \D_{2d} = {(1 + \e^{i3\pi j})(d_{++}\e^{i\pi j/2} + d_{--}\e^{-i\pi j/2}) - 2d_{+-} - 2d_{-+}\e^{i3\pi j} \over (-1+ \e^{i3\pi j})}.  \label{monodromy}
\eea
The terms $\D_{2c}$ and $\D_{2d}$ depend on coefficients $c_{\m\n}$ and $d_{\m\n}$, respectively. 
Although our expression for $\D_1$ here is different from that in Ref. \cite{Siopsis} by a phase factor, our final result is identical to their.

Making use of the formula
\bea
&\;& I_1(\mu,\nu) \equiv  \int_0^{\infty} dz\, z^{-1/2} J_\mu (z) J_\nu (z) = {\sqrt{\pi/2} \G({1 +2\mu +2\nu\over 4}) \over  \G({3-2\mu-2\nu\over 4}) \G({3 + 2\mu -2\nu \over 4} ) \G({3 - 2\mu +2\nu \over 4}) },
\eea
one can obtain explicitly
\bea
&\;& c_{++} = \frac{\left( 3 l^2 + 3 l + 1 - j^2 \right) \,
      \G^2(\frac{1}{4})\, \G(\frac{1-2j}{4})\, \G(\frac{1+2j}{4})\, 
\sin [\frac{\pi(1-2j) }{4}] }{48\,{\pi }^{\frac{3}{2}} \,\sin(\frac{j\,\pi }{2})}; \\
&\;& c_{--} = \frac{\left( 3 l^2 + 3 l + 1 - j^2 \right) \,
      \G^2(\frac{1}{4})\, \G(\frac{1-2j}{4})\, \G(\frac{1+2j}{4})\, 
\sin [\frac{\pi(1+2j) }{4}] }{48\,{\pi }^{\frac{3}{2}} \,\sin(\frac{j\,\pi }{2})}; \\
&\;& c_{+-} = \frac{\left( 3 l^2 + 3 l + 1 - j^2 \right) \,
      \G^2(\frac{1}{4})\, \G(\frac{1-2j}{4})\, \G(\frac{1+2j}{4})\, 
\sin [\frac{\pi(1-2j) }{4}] \sin [\frac{\pi(1+2j) }{4}]}{24 \sqrt{2}\,{\pi }^{\frac{3}{2}} \,\sin(\frac{j\,\pi }{2})}.
\eea
Note that 
\be
c_{--} = -c_{++} (j\to -j); \quad c_{+-} = - c_{-+}(j\to -j). \label{relation}
\ee
These relation are also obeyed by $d_{\m\n}$'s, which can be used to reduce our work. 
With the above results, we are ready to find $\D_1$ and $\D_{2c}$ in eq (\ref{monodromy}):
\bea
&\;& \hskip -1cm \D_1 = - { i(3l^2 + 3l + 1-j^2)\, \G^2(\frac{1}{4})\, \G(\frac{1-2j}{4})\, 
\G(\frac{1+2j}{4})\, \cos({j\pi \over 2})\, \cos(j\pi)
 \over 6 \sqrt{2} \pi^{3/2} [1 + 2\cos(j\pi)] }; \\
&\;& \hskip -1cm \D_{2c} = - { (3l^2 + 3l + 1-j^2)^2\, \G^4(\frac{1}{4})\, 
\G^2(\frac{1-2j}{4})\, \G^2(\frac{1+2j}{4})\, \cos(j\pi) \over 1152 \pi^3 }. \label{D2c}
\eea

The double integral 
\bea
&\;& I_2(\mu_2,\nu_2;\mu_1,\nu_1) \equiv \int_0^{\infty} dz_2 \int_0^{z_2} \, dz_1\, z_2^{-1/2}\, z_1^{-1/2} J_{\mu_2} (z_2) J_{\nu_2} (z_2)  J_{\mu_1} (z_1) J_{\nu_1} (z_1) 
\eea
can be expressed in terms of the generalized hypergeometric functions, but the general formula is quite complicated and not particularly illuminating.  Therefore, we will just give the final result explicitly for the coefficients $d_{++}$ and $d_{+-}$:
\bea
&\;& d_{++} = - \frac{ \pi^2\, \left( 3l^2 + 3l + 1 - j^2 \right)^2\, \cot(\frac{j \pi}{2}) \G(\frac{1}{4})\,
       _{5}G_{4}( \frac{1}{4},\frac{1}{2},\frac{1}{2}, \frac{1+j}{2}, \frac{1-j}{2};
        \frac{5}{4},1, \frac{2+j}{2}, \frac{2-j}{2}; 1)  }{576 \, \sin^2(\frac{j\,\pi }{2})} \nonumber \\
&\;& \hskip 1cm + \frac{{\sqrt{\pi }}\, \left( 3l^2 + 3l + 1 - j^2 \right)^2 \,
     {\cot (\frac{j\,\pi }{2})}\, \cot (j\,\pi )\, \G(\frac{1+2j}{4})\, \G(\frac{1+2j}{2})\, {\G^2(\frac{1+j}{2})} }{288 \sin(\frac{j \pi}{2}) } \nonumber \\ 
&\;& \hskip 1cm _{5}G_{4}(\frac{1}{2},\frac{1+2j}{4}, \frac{1+j}{2},
       \frac{1+j}{2}, \frac{1+2j}{2}; \frac{2+j}{2}, \frac{2+j}{2},\frac{5+2j}{4},1 + j; 1) \nonumber \\
&\;& \hskip 1cm + \frac{\left( 3 l^2 + 3 l + 1 - j^2 \right)^2\,
     {\G^4(\frac{1}{4})}\, {\G^2(\frac{1-2j}{4})}\, {\G^2(\frac{1+2j}{4})}\,
     \sin^2 [\frac{\pi(1-2j) }{4}]\, \sin[\frac{\pi(1+2j)}{4}]}{1152\sqrt{2}\,{\pi }^3 \, \sin^2 (\frac{j\,\pi }{2}) }; \\
&\;& d_{+-} = - \frac{ \pi^2\, \left( 3 l^2 + 3 l + 1 - j^2 \right)^2 \, \G(\frac{1}{4})\,
       _{5}G_{4}( \frac{1}{4},\frac{1}{2},\frac{1}{2}, \frac{1+j}{2}, \frac{1-j}{2};
        \frac{5}{4},1, \frac{2+j}{2}, \frac{2-j}{2}; 1)  }{1152 \, \sin^3(\frac{j\,\pi }{2})} \nonumber \\&\;& \hskip 1cm + \frac{{\sqrt{\pi }}\, \left( 3 l^2 + 3 l + 1 - j^2 \right)^2 \,
     {\cot^2 (\frac{j\,\pi }{2})}\, \cot (j\,\pi )\, \G(\frac{1-2j}{4})\, \G(\frac{1-2j}{2})\, {\G^2(\frac{1-j}{2})} }{576} \nonumber \\ 
&\;& \hskip 1cm _{5}G_{4}(\frac{1}{2},\frac{1-2j}{4}, \frac{1-j}{2},
       \frac{1-j}{2}, \frac{1-2j}{2}; \frac{2-j}{2}, \frac{2-j}{2},\frac{5-2j}{4},1 - j; 1) \nonumber \\
&\;& \hskip 1cm - \frac{\left( 3 l^2 + 3 l + 1 - j^2 \right)^2\,
     {\G^4(\frac{1}{4})}\, {\G^2(\frac{1-2j}{4})}\, {\G^2(\frac{1+2j}{4})}\,
     \sin^2 [\frac{\pi(1-2j) }{4}]\, \sin^2 [\frac{\pi(1+2j)}{4}]}{2304\,{\pi }^3 \, \sin^2 (\frac{j\,\pi }{2}) }.
\eea 
Here, we have used the regularized generalized hypergeometric function $_{5}G_{4}(a_1,a_2,a_3,a_4,a_5; b_1,b_2,b_3,b_4;z)$ so that the pole structure of each term in these expressions are more explicit.  It is related to the usual generalized hypergeometric function by
\be
_{5}G_{4}(a_1,a_2,a_3,a_4,a_5; b_1,b_2,b_3,b_4;z) = {_{5}F_{4}(a_1,a_2,a_3,a_4,a_5; b_1,b_2,b_3,b_4;z) \over \G(b_1) \G(b_2)\G(b_3)\G(b_4)}.
\ee
The other two coefficients can be obtained by relations analogous to those in eq (\ref{relation}) 
\be
d_{--} = -d_{++} (j\to -j); \quad d_{-+} = - d_{+-}(j\to -j).
\ee
On the face of it, each of the $d_{\m\n}$'s has a third order pole coming from terms involving the generalized hypergeometric function when $j$ is an even integer. On closer look, we see there are some cancelation among the divergences and in the end all they have are just simple poles in such limit similar to the $c_{\m\n}$'s.  Another possible divergence arises in $d_{--}$ when $j=1$, which will again be canceled when we calculate the monodromy.

It is now straightforward to obtain $\D_{2d}$ by making use of the following two identities
\bea
&\;&  - 4\pi^{7/2}\, \cos^2(\frac{j \pi}{2})\, \cot(j\pi) \G(\frac{1+2j}{4})\, \G(\frac{1+2j}{2})\, \G^2(\frac{1+j}{2}) \nonumber \\ 
&\;& _{5}G_{4}(\frac{1}{2},\frac{1+2j}{4}, \frac{1+j}{2}, \frac{1+j}{2}, \frac{1+2j}{2}; \frac{2+j}{2}, \frac{2+j}{2},\frac{5+2j}{4},1 + j; 1) \nonumber \\
&\;&  + 4\pi^{7/2}\, \cos^2(\frac{j \pi}{2})\, \cot(j\pi) \G(\frac{1-2j}{4})\, \G(\frac{1-2j}{2})\, \G^2(\frac{1-j}{2}) \nonumber \\ 
&\;& _{5}G_{4}(\frac{1}{2},\frac{1-2j}{4}, \frac{1-j}{2}, \frac{1-j}{2}, \frac{1-2j}{2}; \frac{2-j}{2}, \frac{2-j}{2},\frac{5-2j}{4},1 - j; 1) \nonumber \\ 
&\;& - \G^4(\frac{1}{4})\, \G^2(\frac{1-2j}{4})\, \G^2(\frac{1+2j}{4})\,
     \sin[\frac{\pi(1-2j)}{4}]\, \sin[\frac{\pi(1+2j)}{4}] = 0 ; \\
&\;&  - 4\pi^{7/2}\, \cos(\frac{j \pi}{2})\, \G(\frac{1+2j}{4})\, \G(\frac{1+2j}{2})\, \G^2(\frac{1+j}{2}) \nonumber \\ 
&\;& _{5}G_{4}(\frac{1}{2},\frac{1+2j}{4}, \frac{1+j}{2}, \frac{1+j}{2}, \frac{1+2j}{2}; \frac{2+j}{2}, \frac{2+j}{2},\frac{5+2j}{4},1 + j; 1) \nonumber \\
&\;&  - 4\pi^{7/2}\, \cos(\frac{j \pi}{2})\, \G(\frac{1-2j}{4})\, \G(\frac{1-2j}{2})\, \G^2(\frac{1-j}{2}) \nonumber \\ 
&\;& _{5}G_{4}(\frac{1}{2},\frac{1-2j}{4}, \frac{1-j}{2}, \frac{1-j}{2}, \frac{1-2j}{2}; \frac{2-j}{2}, \frac{2-j}{2},\frac{5-2j}{4},1 - j; 1); \nonumber \\
&\;& + 8\pi^5\, \G(\frac{1}{4})\, 
       _{5}G_{4}( \frac{1}{4},\frac{1}{2},\frac{1}{2}, \frac{1+j}{2}, \frac{1-j}{2};
        \frac{5}{4},1, \frac{2+j}{2}, \frac{2-j}{2}; 1)  \nonumber \\
&\;& - {\G^4(\frac{1}{4})}\, {\G^2(\frac{1-2j}{4})}\, {\G^2(\frac{1+2j}{4})}\,
     \cos({j\pi \over 2})[1 - \cos(j\pi)] = 0. 
\eea 
Eventually, we achieve the following nice result
\bea
&\;& \hskip -1.5cm \D_{2d} = { (3l^2 + 3l + 1-j^2)^2\, \G^4(\frac{1}{4})\, \G^2(\frac{1-2j}{4})\, \G^2(\frac{1+2j}{4})\, \cos(j\pi)
 \over 1152 \pi^3 [1 + 2\cos(j\pi)] },
\eea
where all divergences have been canceled out.

Together with the result from eq (\ref{D2c}), the asymptotic form of quasinormal frequencies of a four dimensional Schwarzschild black hole is found to be 
\bea
&\;& \hskip -1.2cm 4\pi {\om r_0} = (2n+1)\pi i + \ln[1+2\cos(j\pi)] \nonumber \\
&\;& \hskip 0cm - { i(3l^2 + 3l + 1-j^2)\, \G^2(\frac{1}{4})\, \G(\frac{1-2j}{4})\, \G(\frac{1+2j}{4})\, \cos({j\pi \over 2})\, \cos(j\pi)
 \over 6 \sqrt{2} \pi^{3/2} [1 + 2\cos(j\pi)] \sqrt{-\om r_0} }  \nonumber \\
&\;& \hskip 0cm + { (3l^2 + 3l + 1-j^2)^2\, \G^4(\frac{1}{4})\, \G^2(\frac{1-2j}{4})\, 
\G^2(\frac{1+2j}{4})\, \cos^2(j\pi) \over 576 \pi^3 [1 + 2\cos(j\pi)]^2(-\om r_0) } + O[(-\om r_0)^{-3/2}]. \label{qnmformula}
\eea

The physically interested cases are
\bea
&\;& \hskip -2cm {\om_n \over T_H} \approx (2n+1)\pi i +\ln3 + {1-i \over \sqrt{n}} {(l^2 + l - 1) \G^4(1/4) \over 18 \sqrt{2} \pi^{3/2} } +  {i \over n} { (l^2 + l - 1)^2 \G^8(1/4) \over 2592 \pi^3},\;  \hskip 1cm \mbox{for } j=2; \\
&\;& \hskip -2cm {\om_n \over T_H} \approx (2n+1)\pi i +\ln3 + {1-i \over \sqrt{n}} {(l^2 + l + 1/3) \G^4(1/4) \over 6 \sqrt{2} \pi^{3/2} } + {i \over n} { (l^2 + l + 1/3)^2 \G^8(1/4) \over 288 \pi^3},\quad \mbox{for } j=0; \\
&\;& \hskip -2cm {\om_n \over T_H} \approx 2n\pi i +  {i 2\pi(l^2 + l )^2 \over n}, \hskip 8.7cm \mbox{for } j=1. 
\eea
A few comments are in order.  First, all the second order corrections are purely imaginary.
In particular, when $j=2$ (gravitational perturbation) the numerical coefficients of the $i/n$ term (after divided by $4\pi$) are $0.739, 3.58, 49.7$ for $l = 2, 3, 6$, respectively.  They are in good agreement with the known numerical studies \cite{Nollert}.  As for the real part, our result predicts vanishing correction.  For $j=2$, this is again consistent with the numerical results in Ref. \cite{Nollert} for $l = 2, 3$.  For $l=6$ the numerical result is $0.263$, which seems to be contradictory to ours. However, the numerical value for $l=6$ has opposite sign relative to those of $l = 2, 3$.  This is peculiar, since in all other cases a given type of corrections are always of the same sign irrespective of the specific value of angular momentum.  Therefore, we believe more study is needed to clarify whether there is really a discrepancy.  As for the $j=1$ case, the numerical study in Ref. \cite{Cardoso} suggests the leading correction is of the form ${b \over n^{3/2}}$.  However, this does not necessarily mean the two results are inconsistent.  In fact, one can only extract the behavior of the leading correction to the real part from their Fig. 2 and further numerical study is needed to confirm or refute our prediction.

\section{Conclusion}

In sum, we have calculated to second order the correction to the asymptotic form of quasinormal frequencies for Schwarzschild black holes in four dimensions. Most of our results are consistent with the numerical ones when available.  In cases where there seem to be contradiction, we think further numerical studies are needed to clarify the situation.  It would also be helpful if more detailed numerical studies can be carried out for the $j=0$ case so that more thorough comparisons are possible.  It would be interesting to generalize the method to other spacetime backgrounds \cite{Other}.  Extension to higher order is also desirable. It might enable us to find a quantitative prediction for the "algebraically special" frequencies in Schwarzschild black holes, where the quasinormal frequency is purely imaginary and it increases with the fourth power of $l$ \cite{Chandrasekhar,Nollert}.  

\section*{Acknowledgment} 
 
The author thanks Chong-Sun Chu for helpful discussions. The work is supported in part by the National Science Council and the National Center for Theoretical Sciences, Taiwan.

\end{document}